\documentclass[aps,pra,twocolumn,showpacs,preprintnumbers]{revtex4-1}

\usepackage{psfrag,graphicx}
\usepackage{dcolumn}
\usepackage{amsmath,amssymb}
\usepackage{amsfonts}
\usepackage{bm}
\usepackage{amsfonts,amssymb,amsmath}
\usepackage{bbold}
\usepackage{epstopdf}
\usepackage{xcolor}
\usepackage{graphicx}
\usepackage{color}

\usepackage{verbatim}

\newcommand{\jk}[1]{\textcolor{blue}{#1}}

\DeclareSymbolFont{bbold}{U}{bbold}{m}{n}
\DeclareSymbolFontAlphabet{\mathbbold}{bbold}

\usepackage[caption=false]{subfig}

\begin{document}
	
\title{Majorana and parafermion corner states from two coupled sheets of bilayer graphene}
\author{Katharina Laubscher, Daniel Loss, and Jelena Klinovaja}
\affiliation{Department of Physics, University of Basel, Klingelbergstrasse 82, CH-4056 Basel, Switzerland}
	
\begin{abstract}	
We consider a setup consisting of two coupled sheets of bilayer graphene in the regime of strong spin-orbit interaction, where electrostatic confinement is used to create an array of effective quantum wires. We show that for suitable interwire couplings the system supports a topological insulator phase exhibiting Kramers partners of gapless helical edge states, while the additional presence of a small in-plane magnetic field and weak proximity-induced superconductivity leads to the emergence of zero-energy Majorana corner states at all four corners of a rectangular sample, indicating the transition to a second-order topological superconducting phase. 
The presence of strong electron-electron interactions is shown to promote the above phases to their exotic fractional counterparts. In particular, we find that the system supports a fractional topological insulator phase exhibiting fractionally charged gapless edge states and a fractional second-order topological superconducting phase exhibiting zero-energy $\mathbb{Z}_{2m}$ parafermion corner states, where $m$ is an odd integer determined by the position of the chemical potential. 
\end{abstract}
	
\maketitle
	
\section{Introduction}
	
Over the last few decades, topological phases of quantum matter have been the subject of extensive studies, both in theory and in experiments. In particular, a lot of work has been dedicated to the description and classification of topological insulators (TIs) and topological superconductors (TSCs) in various spatial dimensions~\cite{Hasan2010,Qi2011,Sato2017}. Recently, the generalization of conventional TIs and TSCs to \textit{higher-order} TIs and TSCs has attracted strong interest~\cite{Benalcazar2014,Benalcazar2017,Benalcazar2017b,Song2017,Peng2017,Imhof2017,Geier2018,Schindler2018,Hsu2018,Ezawa2018b,Ezawa2018c,Zhu2018,Wang2018,Yan2018,Liu2018,Zhang2018,Wang2018b,Volpez2018,Plekhanov2019,You2018a,You2018b,Laubscher2019,Calugaru2019,Agarwala2019,Yan2019,Franca2019,Zhang2019a,Zhang2019b}. While conventional $d$-dimensional TIs and TSCs exhibit gapless edge states at their $(d-1)$-dimensional boundaries, $n$th-order $d$-dimensional TIs or TSCs exhibit gapless edge states at their $(d-n)$-dimensional boundaries. 

\begin{figure}[!t]
	\centering
	\includegraphics[width=1\columnwidth]{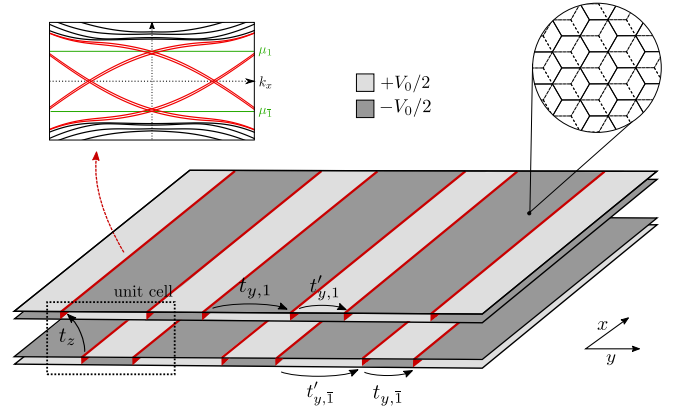}
	\caption{The model consists of two coupled sheets of $AB$-stacked bilayer graphene subject to electrostatic confinement, such that effective 1D wires arise at domain walls between gates set to opposite voltages $\pm V_0/2$. 
	The 
	upper left panel shows the spectrum of an effective wire localized at one of the domain walls with the in-gap states highlighted in red. Note that due to SOI each of the bands is split into two shifted copies. The light green lines labeled $\mu_1$ and $\mu_{\bar 1}$ indicate the values of chemical potential that will be of interest in the remainder of this paper. We now consider an array of such effective wires, where a unit cell is defined as consisting of four wires, see the dashed box. The wires are weakly coupled via a layer-conserving hopping term between neighboring wires within the same unit cell (between neighboring wires belonging to different unit cells) of strength $t_{y,\tau}$ ($t'_{y,\tau}$), as well as via an inter-bilayer hopping term of strength $t_z$. Note that in order to introduce a hierarchy of interwire terms, the wires are arranged in an armchair-like order. In particular, the setup shown here leads to $t_{y,1}\approx t_{y,\bar1}'<t_z<t_{y,1}'\approx t_{y,\bar1}$, as the strength of the hopping terms naturally decreases with the separation of the wires.}
	\label{fig:blg}
\end{figure} 

In the search for suitable platforms to realize topologically non-trivial physics, graphene and graphene-based systems~\cite{CastroNeto2009,DasSarma2011} such as carbon nanotubes and bilayer graphene (BLG) have attracted particular attention. While the unusual low-energy properties of these systems make them interesting in their own right, they also have been proposed to support topologically non-trivial phases of matter, hosting, e.g., gapless edge states or localized Majorana zero modes~\cite{Kane2005,Klinovaja2012c,Sau2013,Egger2012,Kiesel2012,Schaffer2012, Klinovaja2013,Marganska2018,Dutreix2014,Jose2015}. Unfortunately, most of these proposals require strong spin-orbit interaction (SOI) as a crucial ingredient, whereas SOI is weak in standard graphene~\cite{Gmitra2009}. In the last few years, however, considerable experimental progress in creating van der Waals heterostructures has made it possible to induce strong SOI in graphene by proximity to transition metal dichalcogenides (TMDs)~\cite{Kaloni2014,Avsar2014,Gmitra2015,Wang2015,Gmitra2016,Wang2016,Alsharari2016, Kochan2017,Gani2019,Gmitra2017,Khoo2017,Zihlmann2018}, which has led to renewed interest in graphene-based systems as promising candidates to realize topologically non-trivial phases in the laboratory. These considerations, together with the recent interest in higher-order topological phases of matter, have prompted us to devise a graphene-based system realizing second-order topological superconducting phases. In particular, we consider an array of coupled quantum wires arising in bilayer graphene due to electrostatic confinement~\cite{Martin2008,Li2016,Yin2016,Ju2015,Klinovaja2012a}, see Fig.~\ref{fig:blg}. The combined effects of competing interwire hopping terms, an in-plane magnetic field, and proximity-induced superconductivity lead to the formation of zero-energy corner states at all four corners of a rectangular sample. In the non-interacting case, these corner states are Majorana bound states. The major benefit of studying an array of coupled wires, however, is the additional possibility of including the effects of strong electron-electron interactions in an analytically tractable way~\cite{Poilblanc1987,Gorkov1995,Kane2002,Teo2014,Klinovaja2014a,Klinovaja2014b,Sagi2014,Neupert2014,Klinovaja2015,Sagi2015,Meng2015b,Teo2016}. Using  bosonization techniques, we show that suitable interactions can drive the system into a fractional phase exhibiting zero-energy $\mathbb{Z}_{2m}$ parafermion corner states for an odd integer $m$, placing our model in the class of fractional second-order TSCs.

This paper is organized as follows. In Sec.~\ref{sec:model}, we describe a model for our setup, which consists of an array of coupled wires arising in bilayer graphene due to electrostatic confinement. In Sec.~\ref{sec:TI_integer}, we show that this system is a topological insulator for a certain range of parameters. Section~\ref{sec:TI_fractional} extends this result to the interacting case, showing that the system supports fractionally charged edge states for suitable values of chemical potential and sufficiently strong electron-electron interactions. In Sec.~\ref{sec:cornerstates_integer}, we then include suitable superconducting and magnetic perturbations to gap out the helical edge states found previously and show that, in the non-interacting case, the system is driven into a second-order topological superconducting phase with Majorana corner states at all four corners of a rectangular sample. In Sec.~\ref{sec:cornerstates_fractional}, we again extend our analysis to the interacting case and show that the system can be driven into a phase hosting exotic $\mathbb{Z}_{2m}$ parafermion corner states, where $m$ is an odd integer depending on the chemical potential. We summarize our results in Sec.~\ref{sec:conclusions}.


\section{Model}
\label{sec:model}
We consider a system consisting of two coupled sheets of $AB$-stacked bilayer graphene as shown in Fig.~\ref{fig:blg}. Each layer of graphene is a honeycomb lattice consisting of two non-equivalent atoms $A$ and $B$ coupled by a hopping element $t$. Each sheet of bilayer graphene is then composed of two layers of graphene coupled by a hopping amplitude $t_\perp$ between the $A$ atoms of the first layer and the $B$ atoms of the second layer. The effective  Hamiltonian for a single sheet of BLG  in momentum space is then given by
\begin{equation}
\mathcal{H}_\mathrm{BLG}=\hbar v_F(\lambda_z\gamma_xk_x+\gamma_yk_y)+\frac{t_\perp}{2}(\gamma_x\eta_x+\gamma_y\eta_y)-V\eta_z,
\end{equation}
where $\lambda_i$, $\gamma_i$, and $\eta_i$ for $i\in\{x,y,z\}$ are Pauli matrices acting in valley, sublattice, and layer space, respectively, $v_F$ is the Fermi velocity for electrons in graphene, and the term proportional to $V$ describes a potential difference between the two graphene layers. Within each sheet of BLG, electrostatic confinement arising from a spatial modulation of $V$ can then be used to form effective 1D wires. In particular, we consider creating one-dimensional domain walls between gates set to opposite voltages $\pm V_0/2$. This leads to propagating one-dimensional states localized around the region where the voltage changes sign~\cite{Martin2008}. In the following, consider an array of such effective wires, where the wires are arranged in an armchair-like order as shown in Fig.~\ref{fig:blg}. For later convenience, we define a unit cell as consisting of four wires (two belonging to the upper layer, and two belonging to the lower layer). As such, each wire can be labeled by a unit cell index $n$ as well as two indices $(\nu,\tau)$, where $\nu\in\{1,\bar 1\}$ denotes the position within the unit cell and $\tau\in\{1,\bar 1\}$ denotes the layer.

The case of a single effective wire without SOI~\cite{Martin2008} as well as in the presence of curvature-induced SOI~\cite{Klinovaja2012a} has been thoroughly analyzed in previous works. However, even though the curvature-induced SOI is considerably larger than the intrinsic SOI of standard graphene, it is still relatively small~\cite{Ando2000,Hernando2006,Kuemmeth2008,Izumida2009,Chico2009,Klinovaja2011a,Klinovaja2011b,Steele2013}. In order to access a regime with stronger SOI and avoid the need for curvature, we consider a van der Waals heterostructure combining layers of graphene and a TMD~\cite{Kaloni2014,Avsar2014,Gmitra2015,Wang2015,Gmitra2016,Wang2016,Alsharari2016, Kochan2017,Gani2019,Gmitra2017,Khoo2017,Zihlmann2018}. In this case, the proximity-induced SOI is of the form $\mathcal{H}_{so}=\alpha\lambda_z\sigma_z+\alpha_R(\lambda_z\gamma_x\sigma_y-\gamma_y\sigma_x)$, where $\sigma_i$ for $i\in\{x,y,z\}$ is a Pauli matrix acting in spin space~\cite{Gmitra2015,Wang2016}. While the predicted values for $\alpha$ and $\alpha_R$ vary across the literature and depend on the specific TMD that is used, we find that in our case, the Rashba-like term proportional to $\alpha_R$ is suppressed by strong interlayer tunneling $t_\perp$, which is why we focus on the term proportional to $\alpha$~\cite{soi}.

In the following, we consider a step function potential of strength $V_0/2$, where, without loss of generality, we assume $V_0>0$  and focus on the case where the direction of confinement is along the armchair direction of the graphene lattice.  Adapting the results of Refs.~\cite{Klinovaja2012a,Martin2008} to our setup to include the SOI, we find that the bulk gap of the spectrum is given by $V_0$, while there are eight in-gap modes per effective wire, see  the spectrum shown in Fig.~\ref{fig:blg}. Explicitly, the energies of the in-gap modes for wire $\nu$ in the layer $\tau$ are given by
\begin{align}
E_{\lambda\kappa\nu\tau\sigma}&=-\kappa\tau\left[\frac{\hbar v_F\lambda k_x}{2\sqrt{t_\perp}}-\kappa\nu\sqrt{\frac{(\hbar v_Fk_x)^2}{4t_\perp}+\frac{V_0}{2\sqrt{2}}}\,\,\right]^2\nonumber\\&\hspace{5mm}+\kappa\tau\frac{V_0}{\sqrt{2}}+\alpha\lambda\tau\sigma-\mu_\tau.\label{eq:spectrum}
\end{align}
Here, $\lambda\in\{1,\bar 1\}$ is the valley index, $\sigma\in\{1,\bar 1\}$ is the spin projection onto the $z$ axis that is determined by the SOI of strength $\alpha$ (which we take to be of equal magnitude but of opposite sign for the two sheets of BLG), and $\kappa\in\{1,\bar 1\}$ is an additional subband index. For now, we tune the chemical potential in layer $\tau$ to $\mu_\tau=\tau[V_0/(2\sqrt{2})+\alpha]$, while an alternative choice is described in Appendix~\ref{app:mu}. Note that for these values of chemical potential, the sector $\kappa=1$ ($\kappa=\bar 1$) corresponds to modes with small Fermi momenta close to $k_x=0$ (large Fermi momenta far away from $k_x=0$). As an additional simplification, we note that for $\alpha\ll V_0$, the above energy spectrum is approximately linear for small $k_x$ (i.e., in the sector $\kappa=1$) such that
\begin{equation}
E_{\lambda1\nu\tau\sigma}\approx\lambda\nu\tau\hbar v_F \sqrt{\frac{V_0}{2\sqrt{2}t_\perp}}\,k_x+\alpha\tau(\lambda\sigma-1).\label{eq:spectrum_approx}
\end{equation}
To proceed, we work in the regime of strong spin-orbit interaction, which allows us to linearize the spectrum of each channel around the respective Fermi point. 
The electron operator for the $n$th unit cell can then be represented as $\Psi_n=\sum_{\kappa,\nu,\tau,\sigma}(R_{n\kappa\nu\tau\sigma}e^{ik_F^{(\nu\tau)\kappa\nu\tau\sigma}x}+L_{n\kappa\nu\tau\sigma}e^{ik_F^{(\bar{\nu\tau})\kappa\nu\tau\sigma}x})$, where $R_{n\kappa\nu\tau\sigma}(x)$ [$L_{n\kappa\nu\tau\sigma}(x)$] are slowly right-moving [left-moving] fields and $k_F^{\lambda\kappa\nu\tau\sigma}$ are the respective Fermi momenta. Note that here and in the following we implicitly assume the presence of weak but finite intervalley scattering, which is why we do no longer consider the valley degree of freedom to be a good quantum number but treat modes differing only in their valley index as right and left moving modes of the same species. The effective Hamiltonian describing the uncoupled wires can be written as
%
%
\begin{align}
\label{eq:H0lin}
H_0&=-i\hbar\sum_{n,\kappa,\nu,\tau,\sigma}\int dx\,v_{\kappa}(R_{n\kappa\nu\tau\sigma}^\dagger\partial_xR_{n\kappa\nu\tau\sigma}\nonumber\\&\hspace{37mm}- L_{n\kappa\nu\tau\sigma}^\dagger\partial_xL_{n\kappa\nu\tau\sigma}),
\end{align}
where we have made use of the fact that the Fermi velocities 
of the different branches for a fixed $\kappa$ are approximately the same given that $\alpha\ll V_0$. Explicitly, we note from Eq.~(\ref{eq:spectrum_approx}) that, for the values of chemical potential of interest to us, we approximately have $v_1=v_F\sqrt{V_0/(2\sqrt{2}t_\perp)}$.

We will now couple neighboring wires in various ways. Neglecting all fast oscillating terms, we consider the interwire Hamiltonian $H_\perp=H_y+H_y'+H_z$, where
\begin{align}
&H_y=\sum_{n,\nu,\tau} t_{y,\tau}\int dx\, R_{n1\nu\tau(\nu\tau)}^\dagger L_{n1\bar\nu\tau(\nu\tau)} +\mathrm{H.c.},\label{eq:Hylin}\\
&H_y'=\sum_{n,\tau} t_{y,\tau}'\int dx\, [R_{(n+1)11\tau\tau}^\dagger L_{n1\bar1\tau\tau} \nonumber\\&\hspace{30mm}+L_{(n+1)11\tau\bar\tau}^\dagger R_{n1\bar1\tau\bar\tau}]+\mathrm{H.c.},
\label{eq:Hy'lin}\\
&H_z=\sum_{n,\kappa,\nu,\tau,\sigma} t_z\int dx\,R^\dagger_{n\kappa\nu\tau\sigma}L_{n\kappa\nu\bar\tau\sigma}+\mathrm{H.c.}, \label{eq:Hzlin}
\end{align}
with $0\leq t_{y,\tau},t_{y,\tau}',t_z\ll \alpha$. Here, $t_{y,\tau}$ ($t_{y,\tau}'$) is a spin-conserving intralayer hopping element between neighboring wires within the same unit cell (between neighboring wires belonging to different unit cells) and $t_z$ is a spin-conserving hopping element between neighboring wires belonging to different layers. 
The strength of these hopping amplitudes can be controlled by varying the interwire distance as well as the strength and the shape of the confinement potential.

Furthermore, if a superconducting TMD such as $\mathrm{NbSe_2}$ is used, superconductivity will be induced in the graphene bilayers~\cite{Gani2019}. The corresponding effective Hamiltonian then reads
\begin{equation}
\label{eq:Hsc}
H_{sc}=\Delta_{sc}\sum_{n,\kappa,\nu,\tau,\sigma}\sigma\int dx\,  R_{n\kappa\nu\tau\sigma}^\dagger L_{n\kappa\nu\tau\bar\sigma}^\dagger+\mathrm{H.c.} 
\end{equation}
Additionally, we consider the effect of an in-plane Zeeman field along the $x$ direction. Combined with intervalley scattering, which we assume to be present in the system with broken translational invariance, this term takes the form
\begin{equation}
\label{eq:HZ}
H_{Z}=\Delta_Z\sum_{n,\nu,\tau}\int dx\,R_{n1\nu\tau(\nu\tau)}^\dagger L_{n1\nu\tau(\bar{\nu\tau})}+\mathrm{H.c}.
\end{equation}
Finally, the total Hamiltonian is defined as $H=H_0+H_\perp+H_Z+H_{sc}$. In the remainder of this paper, we will focus on the regime $\Delta_{sc}$, $\Delta_Z\ll t_{y,\tau}, t_{y,\tau}', t_z$ such that the superconducting and Zeeman term can be treated as weak perturbations to the interwire terms. Numerically, however, our analysis can be extended to the non-perturbative regime, confirming that the found topological properties persist as long as the bulk gap is not closed.

\begin{figure}[b]
	\includegraphics[width=\columnwidth]{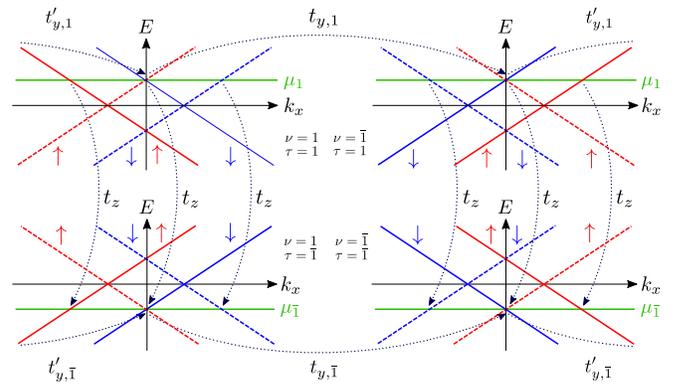}
	\caption{The spectrum of a single unit cell in the non-interacting case, where the chemical potential in layer $\tau$ is tuned to $\mu_\tau=\tau[V_0/(2\sqrt{2})+\alpha]$. For simplicity, only the sector $\kappa=1$ is shown. The branches $R_{n1\nu\tau(\bar{\nu\tau})}$ and $L_{n1\nu\tau(\nu\tau)}$ are trivially gapped by interlayer hopping of strength $t_z$. For the branches $R_{n1\nu\tau(\nu\tau)}$ and $L_{n1\nu\tau(\bar{\nu\tau})}$, on the other hand, the interlayer hopping term competes with intracell hopping of strength $t_{y,\tau}$ and intercell hopping of strength $t_{y,\tau}'$.}
	\label{fig:spectrum_integer}
\end{figure} 

\section{Topological insulator phase}
\label{sec:TI}

\subsection{Non-interacting case}
\label{sec:TI_integer}
In this section, we demonstrate that our model supports a TI phase with Kramers partners of gapless edge states propagating along the edges of a large but finite sample. For this, we set $\Delta_Z=\Delta_{sc}=0$. As can immediately be verified from Eqs.~(\ref{eq:Hylin})-(\ref{eq:Hzlin}), the branches with $\kappa=\bar 1$ are trivially gapped by interlayer hopping. As such, we focus on the sector $\kappa=1$ in the following. For this sector, we find that the branches $R_{n1\nu\tau(\bar{\nu\tau})}$ and $L_{n1\nu\tau(\nu\tau)}$ are trivially gapped by interlayer hopping, whereas the different hopping processes compete for the branches $R_{n1\nu\tau(\nu\tau)}$ and $L_{n1\nu\tau(\bar{\nu\tau})}$, see Fig.~\ref{fig:spectrum_integer}. In the following, we are interested in the regime $t_{y,1}\approx t_{y,\bar1}'<t_z<t_{y,1}'\approx t_{y,\bar1}$. Such a hierarchy is natural for an armchair-like arrangement of the effective wires as shown in Fig.~\ref{fig:blg}, as the strength of the hopping terms can be expected to decrease with the separation of the wires. For simplicity, let us assume that $t_{y,1}=t_{y,\bar1}'=0$ and $t_{y,1}'=t_{y,\bar1}$. By direct inspection of Eqs.~(\ref{eq:Hylin})-(\ref{eq:Hzlin}), we find that the bulk of the system is fully gapped.

In order to find edge states in a system that is finite along the $y$ direction and consists of $N$ unit cells, we note that the parameter regime of our interest is in the same part of the topological phase diagram as the regime $t_z\ll t_{y,1}'$, see Appendix~\ref{app:phasediagram}. In this limit, we find that the two modes $R_{11111}$ and $L_{1111\bar1}$ ($R_{N1\bar11\bar1}$ and $L_{N1\bar111}$) at the left (right) edge of the system stay gapless. The presence of these helical edge states is not affected by deviations from the above fine-tuned point as long as the bulk gap does not close.

Let us now assume that the system is finite along the $x$ direction and infinite along the $y$ direction. We apply the standard procedure of matching decaying eigenfunctions to find edge states propagating along the $y$ direction~\cite{Klinovaja2012b}. 
The projection of the Hamiltonian $H=H_0+H_\perp$ onto the sector $\kappa=1$ can then be written in momentum space as $H=\sum_{k_y}\int dx\, \Psi_{k_y}^\dagger \mathcal H(k_y)\Psi_{k_y}$, with the Hamiltonian density $\mathcal{H}(k_y)$ given by
\begin{widetext}
	\begin{equation}
	\begin{split}
	\mathcal{H}(k_y)&=-i\hbar v_1\partial_x\rho_z+\{[t_{y,1}+t_{y,1}'\mathrm{cos}(k_ya_y)](1+\tau_z)+[t_{y,\bar1}+t_{y,\bar1}'\mathrm{cos}(k_ya_y)](1-\tau_z)\}(\nu_x\rho_x-\nu_y\tau_z\sigma_z\rho_y)/4\\&\hspace{5mm}-[t_{y,1}'\mathrm{sin}(k_ya_y)(1+\tau_z)+t_{y,\bar1}'\mathrm{sin}(k_ya_y)(1-\tau_z)](\nu_y\rho_x+\nu_x\tau_z\sigma_z\rho_y)/4+t_z\tau_x\rho_x\label{eq:Hky}
	\end{split}
	\end{equation}
\end{widetext}
in the basis $\Psi_{k_y}$= ($R_{k_y1111}$, $L_{k_y1111}$, $R_{k_y111\bar1}$, $L_{k_y111\bar1}$, $R_{k_y11\bar11}$, $L_{k_y11\bar11}$, $R_{k_y11\bar1\bar1}$, $L_{k_y11\bar1\bar1}$, $R_{k_y1\bar111}$, $L_{k_y1\bar111}$, $R_{k_y1\bar11\bar1}$, $L_{k_y1\bar11\bar1}$, $R_{k_y1\bar1\bar11}$, $L_{k_y1\bar1\bar11}$, $R_{k_y1\bar1\bar1\bar1}$, $L_{k_y1\bar1\bar1\bar1}$).
Here, $\nu_{i}$, $\tau_i$, $\sigma_i$, and $\rho_i$ for $i\in\{x,y,z\}$ are Pauli matrices acting in wire, layer, spin, and right/left mover space, respectively, and $a_y$ is the size of a unit cell in the $y$ direction. Next, we focus on $k_y=0$ and a single edge of the system at $x=0$. In order to satisfy vanishing boundary conditions, we require $R_{k_y1\nu\tau\sigma}(0)=-L_{k_y1\nu\tau\sigma}(0)$. From this condition, we find that, given $t_{y,\bar1}>t_z$, there are two exponentially decaying solutions localized to the edge of the system. These are given by
\begin{equation}
\begin{aligned}
\label{eq:ymodes}
\Phi_+&=\left(-a,b,0,0,ib,-ia,0,0,-ib,ia,0,0,a,-b,0,0\right)^T,\\
\Phi_-&=\left(0,0,b,-a,0,0,ia,-ib,0,0,-ia,ib,0,0,-b,a\right)^T,
\end{aligned}
\end{equation}
in the basis of $\Psi_{k_y=0}$. Here we defined $a=e^{-x/\xi_1}$ and $b=e^{-x/\xi_2}$ with $\xi_1=\hbar v_1/(t_{y,\bar1}-t_z)$ and $\xi_2=\hbar v_1/t_z$. It is straightforward to verify that these edge states are Kramers partners and related by time-reversal via $\Phi_-=-i\sigma_y\rho_x\mathcal{K}\Phi_+$, where $\mathcal{K}$ denotes the complex conjugation.

Putting together all of the above results, we conclude that our system is in a topological insulator phase with a Kramers pair of gapless edge states running along the edges of a large but finite sample. 

\subsection{Interacting case}
\label{sec:TI_fractional}
Let us now address the construction of the fractional counterpart of the above phase.
For this, we tune the chemical potential in layer $\tau$ to $\mu_\tau=\tau[V_0/(2\sqrt{2})+\alpha/m]$, where $m$ is an odd integer and $m=1$ reproduces the non-interacting case discussed above. Again, the interlayer hopping term given in Eq.~(\ref{eq:Hzlin}) trivially gaps out the sector $\kappa=\bar 1$ corresponding to large Fermi momenta. Therefore, we again focus on the sector $\kappa=1$.

\begin{figure}[]
	\includegraphics[width=\columnwidth]{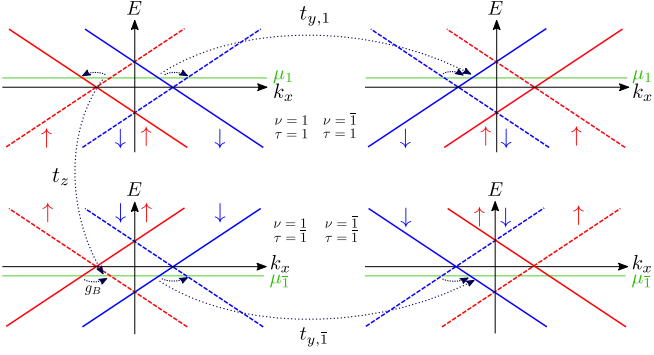}
	\caption{The spectrum of a single unit cell for $\mu_\tau=\tau[V_0/(2\sqrt{2})+\alpha/3]$, corresponding to $m = 3$. Again, only the sector $\kappa=1$ is shown. Both inter- and intracell hopping do not conserve momentum unless they are dressed by backscattering events of strength $g_B$ arising from strong electron-electron interactions. In order to commute with the other interwire terms, the interlayer hopping term has to be dressed by electron-electron interactions as well. For clarity of the presentation, not all the terms of the interwire Hamiltonian $H_\perp$ [see Eqs.~(\ref{eq:Hy_dressed})-(\ref{eq:Hz_dressed})] are shown.}
	\label{fig:spectrum_fractional}
\end{figure}

As a first step, we note that for $m>1$ the hopping processes between neighboring wires belonging to the same layer [see Eqs.~(\ref{eq:Hylin})  and (\ref{eq:Hy'lin})] no longer conserve momentum. However, momentum-conserving terms can be constructed by including single-electron backscattering processes arising from strong electron-electron interactions, see Fig.~\ref{fig:spectrum_fractional} for a graphical illustration in the case $m=3$. Explicitly, the dressed interwire terms are given by
\begin{widetext}
	\begin{align}
	&H_y^{(m)}=\sum_{n,\nu,\tau} t_{y,\tau}^{(m)}\int dx\, (R_{n1\nu\tau(\nu\tau)}^\dagger L_{n1\nu\tau(\nu\tau)})^{\frac{m-1}{2}} R_{n1\nu\tau(\nu\tau)}^\dagger L_{n1\bar\nu\tau(\nu\tau)} (R_{n1\bar\nu\tau(\nu\tau)}^\dagger L_{n1\bar\nu\tau(\nu\tau)})^{\frac{m-1}{2}}+\mathrm{H.c.},\label{eq:Hy_dressed}\\
	&H_y'^{(m)}=\sum_{n,\tau} t_{y,\tau}'^{(m)}\int dx\, \Big[(R_{(n+1)11\tau\tau}^\dagger L_{(n+1)11\tau\tau})^{\frac{m-1}{2}} R_{(n+1)11\tau\tau}^\dagger L_{n1\bar1\tau\tau} (R_{n1\bar1\tau\tau}^\dagger L_{n1\bar1\tau\tau})^{\frac{m-1}{2}}\nonumber\\&\hspace{39mm}+(L_{(n+1)11\tau\bar\tau}^\dagger R_{(n+1)11\tau\bar\tau})^{\frac{m-1}{2}}L_{(n+1)11\tau\bar\tau}^\dagger R_{n1\bar1\tau\bar\tau} (L_{n1\bar1\tau\bar\tau}^\dagger R_{n1\bar1\tau\bar\tau})^{\frac{m-1}{2}}\Big]+\mathrm{H.c.}\label{eq:Hy'_dressed}
	\end{align}
\end{widetext}
Here, $t_{y,\tau}^{(m)}\propto t_{y,\tau}g_B^{m-1}$ and $t_{y,\tau}'^{(m)}\propto t_{y,\tau}'g_B^{m-1}$, where $g_B$ is the amplitude of a single-electron backscattering process. In the following, let us assume that the above terms flow to strong coupling in a renormalization group (RG) sense. This can always be achieved if their bare coupling constants are sufficiently large or if their scaling dimensions are the lowest ones among all possible competing terms. The original interlayer hopping term given in Eq.~(\ref{eq:Hzlin}) does not commute with the above terms and therefore cannot order simultaneously. Instead, the interlayer term that commutes with the above terms is, to lowest order, given by the dressed term
\begin{align}
H_z^{(m)}&=\sum_{n,\nu,\tau,\sigma} t_z^{(m)}\int dx\,(R^\dagger_{n1\nu\tau\sigma}L_{n1\nu\tau\sigma})^{\frac{m-1}{2}}\label{eq:Hz_dressed}\\&\hspace{10mm}\times R^\dagger_{n1\nu\tau\sigma}L_{n1\nu\bar\tau\sigma}(R_{n1\nu\bar\tau\sigma}^\dagger L_{n1\nu\bar\tau\sigma})^{\frac{m-1}{2}}+\mathrm{H.c.},\nonumber
\end{align}
where again $t_{z}^{(m)}\propto t_{z}g_B^{m-1}$. Following the same arguments as above, we assume that this term flows to strong coupling. The total interwire Hamiltonian in the interacting case is now defined as $H_\perp^{(m)}=H_y^{(m)}+H_y'^{(m)}+H_z^{(m)}$.

In order to facilitate the analytical description of the interacting system, we introduce bosonic fields $\phi_{1n1\nu\tau\sigma}(x)$ and $\phi_{\bar1n1\nu\tau\sigma}(x)$ defined via $R_{n1\nu\tau\sigma}(x)=e^{i\phi_{1n1\nu\tau\sigma}(x)}$ and $L_{n1\nu\tau\sigma}(x)=e^{i\phi_{\bar1n1\nu\tau\sigma}(x)}$. The fields $\phi_{rn1\nu\tau\sigma}(x)$ satisfy the non-local commutation relation $[\phi_{rn1\nu\tau\sigma}(x),\phi_{r'n'1\nu'\tau'\sigma'}(x')]=ir\pi \delta_{rr'}\delta_{nn'}\delta_{\nu\nu'}\delta_{\tau\tau'}\delta_{\sigma\sigma'}\mathrm{sgn}(x-x')$. With this choice, $ R_{n1\nu\tau\sigma}$ and $L_{n1\nu\tau\sigma}$ satisfy the proper fermionic anticommutation relations among themselves, while the commutation relations between different species can be satisfied by an appropriate choice of Klein factors~\cite{Giamarchi2004}, which we will not explicitly include here. The dressed interwire terms given in Eqs.~(\ref{eq:Hy_dressed})-(\ref{eq:Hz_dressed}) can be simplified by introducing new bosonic operators $\eta_{rn1\nu\tau\sigma}(x)=\frac{m+1}{2}\phi_{rn1\nu\tau\sigma}(x)-\frac{m-1}{2}\phi_{\bar{r}n1\nu\tau\sigma}(x)$. The new fields obey the commutation relations
$[\eta_{rn1\nu\tau\sigma}(x),\eta_{r'n'1\nu'\tau'\sigma'}(x')]=irm\pi\delta_{rr'}\delta_{nn'}\delta_{\nu\nu'}\delta_{\tau\tau'}\delta_{\sigma\sigma'}\mathrm{sgn}(x-x')$ and, for $m >1$, they carry fractional charge $e/m$~\cite{Teo2014}. In terms of these new fields, the dressed interwire terms take the form
\begin{align}
&H_y^{(m)}=2\sum_{n,\nu,\tau} t_{y,\tau}^{(m)}\int dx\, \mathrm{cos}(\eta_{1n1\nu\tau(\nu\tau)}-\eta_{\bar1n1\bar\nu\tau(\nu\tau)}),\\
&H_y'^{(m)}=2\sum_{n,\tau} t_{y,\tau}'^{(m)}\int dx\, [\mathrm{cos}(\eta_{1(n+1)11\tau\tau}-\eta_{\bar1n1\bar1\tau\tau})\nonumber\\&\hspace{30mm}+\mathrm{cos}(\eta_{\bar1(n+1)11\tau\bar\tau}-\eta_{1n1\bar1\tau\bar\tau})],\\
&H_z^{(m)}=2\sum_{n,\nu,\tau,\sigma} t_z^{(m)}\int dx\,\mathrm{cos}(\eta_{1n1\nu\tau\sigma}-\eta_{\bar1n1\nu\bar\tau\sigma}).
\end{align}
We note that the bulk of the system is now fully gapped, while for a system consisting of $N$ unit cells the modes $\eta_{111111}$ and $\eta_{\bar11111\bar1}$ ($\eta_{1N1\bar11\bar1}$ and $\eta_{\bar1N1\bar111}$) at the left edge (right edge) of the system stay gapless. These edge states carry fractional charges $e/m$, as expected for a fractional TI.

In order to study the emerging fractional edge states further, let us define new composite chiral fermion operators
$R^{(m)}_{n1\nu\tau\sigma}=e^{i\eta_{1n1\nu\tau\sigma}}$ and $ L^{(m)}_{n1\nu\tau\sigma}= e^{i\eta_{\bar1n1\nu\tau\sigma}}$.
In terms of these new composite fields, the dressed interwire terms simplify to
\begin{align}
&H_y^{(m)}=\sum_{n,\nu,\tau} t_{y,\tau}^{(m)}\int dx\, R_{n1\nu\tau(\nu\tau)}^{(m)\dagger} L_{n1\bar\nu\tau(\nu\tau)}^{(m)} +\mathrm{H.c.},\label{eq:Hylin_int}\\
&H_y'^{(m)}=\sum_{n,\tau} t_{y,\tau}'^{(m)}\int dx\, [R_{(n+1)11\tau\tau}^{(m)\dagger} L_{n1\bar1\tau\tau}^{(m)} \nonumber\\&\hspace{33mm}+L_{(n+1)11\tau\bar\tau}^{(m)\dagger} R_{n1\bar1\tau\bar\tau}^{(m)}]+\mathrm{H.c.},
\label{eq:Hy'lin_int}\\
&H_z^{(m)}=\sum_{n,\nu,\tau,\sigma} t_z^{(m)}\int dx\,R_{n1\nu\tau\sigma}^{(m)\dagger}L_{n1\nu\bar\tau\sigma}^{(m)}+\mathrm{H.c.},\label{eq:Hzlin_int}
\end{align}
from which we recover the non-interacting case for $m=1$. Indeed, $H_\perp^{(m)}$ has the exact same form as in the non-interacting case, except that $R_{n1\nu\tau\sigma}$ ($L_{n1\nu\tau\sigma}$) is replaced by $R_{n1\nu\tau\sigma}^{(m)}$  ($L_{n1\nu\tau\sigma}^{(m)}$). We can now repeat the analysis from the non-interacting case for the new fields $R_{n1\nu\tau\sigma}^{(m)}$, $L_{n1\nu\tau\sigma}^{(m)}$ to find that the branches $R^{(m)}_{n1\nu\tau(\bar{\nu\tau})}$ and $L^{(m)}_{n1\nu\tau(\nu\tau)}$ are again fully gapped, while the branches $R^{(m)}_{n1\nu\tau(\nu\tau)}$ and $L^{(m)}_{n1\nu\tau(\bar{\nu\tau})}$ yield two gapless modes $R^{(m)}_{11111}$ and $L^{(m)}_{1111\bar1}$ ($R^{(m)}_{N1\bar11\bar1}$ and $L^{(m)}_{N1\bar111}$) at the left (right) edge  of the system.

As in the non-interacting case, we also consider a semi-infinite geometry where the system is finite along the $x$ direction and infinite along the $y$ direction. By introducing the Fourier transforms $R_{k_y1\nu\tau\sigma}^{(m)}$ and $L_{k_y1\nu\tau\sigma}^{(m)}$ of the composite fields, one can repeat the procedure of matching decaying eigenfunctions employed in the non-interacting case and obtain analogous expressions for the gapless edge states propagating along the $y$ direction, which are now given by
\begin{equation}
\begin{aligned}
\label{eq:ymodes_int}
\Phi_+&=\left(-\tilde{a},\tilde{b},0,0,i\tilde{b},-i\tilde{a},0,0,-i\tilde{b},i\tilde{a},0,0,\tilde{a},-\tilde{b},0,0\right)^T,\\
\Phi_-&=\left(0,0,\tilde{b},-\tilde{a},0,0,i\tilde{a},-i\tilde{b},0,0,-i\tilde{a},i\tilde{b},0,0,-\tilde{b},\tilde{a}\right)^T,
\end{aligned}
\end{equation}
in the basis $\Psi_{k_y=0}^{(m)}$, which corresponds to $\Psi_{k_y=0}$ but with $R_{n1\nu\tau\sigma}$ ($L_{n1\nu\tau\sigma}$) replaced by $R_{n1\nu\tau\sigma}^{(m)}$  ($L_{n1\nu\tau\sigma}^{(m)}$). Furthermore, we have defined $\tilde{a}=e^{-x/\xi_1^{(m)}}$ and $\tilde{b}=e^{-x/\xi_2^{(m)}}$ for $\xi_1^{(m)}=\hbar v_1^{(m)}/(t_{y,\bar1}^{(m)}-t_z^{(m)})$ and  $\xi_2^{(m)}=\hbar v_1^{(m)}/t_z^{(m)}$, where $v_1^{(m)}$ is the velocity of the composite fields.

By continuity, we therefore find that our system hosts a Kramers pair of fractionally charged gapless edge states running along the edges of a large but finite sample, which allows us to identify our system as a fractional topological insulator. This means that we can write an effective edge theory in terms of two conjugate bosonic fields $\eta_1$ and $\eta_{\bar{1}}$ with $[\eta_{r}(l),\eta_{r'}(l')]=irm\pi\delta_{rr'}\mathrm{sgn}(l-l')$, where $l$ is an edge coordinate which is defined mod $2[L+(N-1)a_y]$ and runs along the edge of the sample in the counterclockwise direction~\cite{Yan2018}.

\section{Majorana and parafermion corner states}
\label{sec:cornerstates}

\subsection{Non-interacting case}
\label{sec:cornerstates_integer}

\begin{figure}[tb]
	\centering
	\includegraphics[width=0.8\columnwidth]{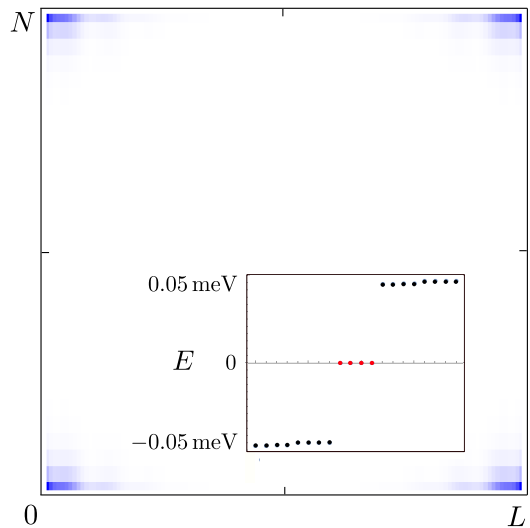}
	\caption{Probability density of low-energy states obtained numerically for a system of $N=30$ unit cells consisting of four effective wires of length $L=1.2\,\mathrm{\mu m}$ each. The single-wire spectrum is obtained by discretizing Eq.~(\ref{eq:spectrum_approx}) and by setting $V_0=100\,\mathrm{meV}$, $t=2.7\,\mathrm{eV}$, $t_\perp=0.34\,\mathrm{eV}$, and $\alpha=1\,\mathrm{meV}$. The other parameters [see Eqs.~(\ref{eq:Hylin})-(\ref{eq:HZ})] are chosen as $t_{y,1}=0.01\,\mathrm{meV}$, $t_{y,1}'=0.79\,\mathrm{meV}$, $t_{y,\bar1}=0.81\,\mathrm{meV}$, $t_{y,\bar1}'=0.02\,\mathrm{meV}$, $t_z=0.4\,\mathrm{meV}$, $\Delta_Z=0.4\,\mathrm{meV}$, and $\Delta_{sc}=0.05\,\mathrm{meV}$. One Majorana corner state is localized at each of the four corners of the system. The inset demonstrates that the energies of these states (red dots) are indeed at zero.}
	\label{fig:majoranas}
\end{figure}

In this section, we show that the terms $H_{sc}$ and $H_Z$ [see Eqs.~(\ref{eq:Hsc}) and (\ref{eq:HZ})] can drive the system into a second-order topological superconducting phase. Again, we start by treating the non-interacting case. Importantly, we consider $\Delta_{sc}$ and $\Delta_Z$ to be small enough not to modify the bulk gap structure. However, they may modify the low-energy behavior of the system by gapping out the helical edge states found above. This statement is confirmed explicitly by considering the effective low-energy edge theory. We assume that the system size is sufficiently large such that far away from the corners all four edges can be treated independently. Crucially, we find that the Zeeman term $\mathcal{H}_Z=\Delta_Z(\sigma_x\rho_x-\nu_z\tau_z\sigma_y\rho_y)$ does not open a gap at $k_y=0$ in the spectrum of the edges states propagating along the $y$ direction. This can be verified explicitly by using the form of the edge state wave functions given in Eq.~(\ref{eq:ymodes}), for which we find $\langle\Phi_+|\mathcal{H}_Z|\Phi_-\rangle=\langle\Phi_+|\mathcal{H}_Z|\Phi_+\rangle=\langle\Phi_-|\mathcal{H}_Z|\Phi_-\rangle=0$. 
Alternatively, one can arrive at the same conclusion in a more general way by exploiting the symmetries of the system. Indeed, at $k_y=0$ the system has an additional symmetry represented by the operator $\mathcal{O}=\nu_z\tau_y\sigma_z\rho_x$, which anticommutes with the Hamiltonian $\mathcal{H}(k_y=0)$ [see Eq.~(\ref{eq:Hky})] for $t_{y,1}+t_{y,1}'=t_{y,\bar1}+t'_{y,\bar1}$. In addition,  $\Phi_\pm$ defined in Eq.~(\ref{eq:ymodes}) are eigenstates of $\mathcal{O}$: $\mathcal{O}\Phi_\pm=\Phi_\pm$. Furthermore, we find 
$\{\mathcal{H}_Z,\mathcal{O}\}=0$, which then implies $\langle\Phi_+|\mathcal{H}_Z|\Phi_-\rangle=0$. 
%
All other matrix elements are trivially zero as the edge states $\Phi_\pm$ are eigenstates of $\sigma_z$, showing that the magnetic term indeed does  not open a gap in the spectrum of the edge states propagating along the $y$ direction. Therefore,  these edge states can only be gapped by superconductivity, whereas in the spectrum of the edge states propagating along the $x$ direction both mechanisms can in principle open a gap. If we choose $|\Delta_Z|>|\Delta_{sc}|$, the magnetic term dominates over the superconducting one such that it is responsible for gapping  the edge states propagating along the $x$ direction. In analogy to previous works studying domain walls between competing gapping mechanisms in systems with helical edge states~\cite{Motruk2013,Clarke2013,Cheng2012}, we find localized Majorana zero modes at the domain walls between the regions where the superconducting/magnetic term dominates, which in this case means at all four corners of the system. However, in contrast to previous works, we apply both the superconducting as well as the magnetic term throughout the entire system. Figure~\ref{fig:majoranas} verifies our results numerically. Importantly, our numerical analysis confirms that the corner states are robust against small deviations from the fine-tuned point $t_{y,1}+t_{y,1}'=t_{y,\bar1}+t'_{y,\bar1}$. In addition, we confirmed numerically that the zero-energy corner states are robust against disorder that breaks all spatial symmetries but neither closes the bulk nor the edge state gaps. This confirms that the Majorana corner states are protected purely by the particle-hole symmetry enforced by superconductivity, while the spatial symmetry $\mathcal{O}$ is not playing a crucial role.

\subsection{Interacting case}
\label{sec:cornerstates_fractional}

The above results can be extended rather straightforwardly to the interacting case. In order to gap out the edge states given in Eq.~(\ref{eq:ymodes_int}), the Zeeman term as well as the superconducting term need to be dressed by interactions in the standard way. To lowest order, the terms that can open a gap in the edge state spectrum are given by
\begin{align}
H_{sc}^{(m)}&=\Delta_{sc}^{(m)}\sum_{n,\nu,\tau,\sigma}\sigma\int dx\,  R_{n1\nu\tau\sigma}^{(m)\dagger} L_{n1\nu\tau\bar\sigma}^{(m)\dagger}+\mathrm{H.c.},\label{eq:Hsclin_int}\\
H_{Z}^{(m)}&=\Delta_Z^{(m)}\sum_{n,\nu,\tau}\int dx\,R_{n1\nu\tau(\nu\tau)}^{(m)\dagger} L_{n1\nu\tau(\bar{\nu\tau})}^{(m)}+\mathrm{H.c}.\label{eq:HZlin_int},
\end{align}
with $\Delta_{sc}^{(m)}\propto\Delta_{sc}g_B^{m-1}$ and $\Delta_Z^{(m)}\propto \Delta_Zg_B^{m-1}$. Again, these terms have exactly the same form as in the non-interacting case, except that $R_{n1\nu\tau\sigma}$ ($L_{n1\nu\tau\sigma}$) is replaced by $R_{n1\nu\tau\sigma}^{(m)}$  ($L_{n1\nu\tau\sigma}^{(m)}$). In the following, we assume that the above terms are substantially smaller than the interwire terms and, therefore, will not modify the bulk gap structure but may open a gap in the spectrum of edge states. Starting from the fractional topological insulator phase established in Sec.~\ref{sec:TI_fractional}, we note that the projection of the above terms onto the gapless edge states commutes with the interwire terms and therefore these terms can order simultaneously.
  In particular, the above terms can again always be made relevant if their bare coupling constants are of order unity or if their scaling dimensions are the lowest ones among all competing terms~\cite{Kane2002,Teo2014}.

We can now repeat the above symmetry argument to find that the magnetic term does not open a gap in the spectrum of the edge states propagating along the $y$ direction; this gap is opened by superconductivity only. On the other hand, the edge states propagating along the $x$ direction are gapped by the Zeeman term for $|\Delta_Z^{(m)}|>|\Delta_{sc}^{(m)}|$. We are thus effectively dealing with domain walls occurring naturally at the corners of a fractional 2D TI, 
despite the fact that
the superconducting and magnetic terms are uniform and act both simultaneously on the entire system.
Given this analogy to domain walls, we  can follow  Refs.~\cite{Cheng2012,Clarke2013,Lindner2012,Motruk2013} and show that every domain wall between a region gapped by superconductivity and a region gapped by a magnetic field hosts a zero-energy parafermion bound state that is spatially localized to the domain wall, i.e., to the corner of the sample. 
To make this statement explicit in terms of the fields considered here, we rewrite the left and right moving fields $\eta_{1}$ and $\eta_{\bar 1}$ describing the low-energy edge theory in terms of conjugate fields $\varphi=(\eta_1-\eta_{\bar 1})/(2m)$ and $\theta=(\eta_1+\eta_{\bar 1})/(2m)$ with $[\varphi(l),\theta(l')]=\frac{i\pi}{2m}\mathrm{sgn}(l-l')$. The dressed superconducting and magnetic terms given in Eqs.~(\ref{eq:Hsclin_int}) and (\ref{eq:HZlin_int}) projected onto the low-energy part of the spectrum now take the form $H_{sc}^{(m)}\propto\Delta_{sc}^{(m)}\int dl\,\mathrm{cos}(2m\theta)$ and $ H_{Z}^{(m)}\propto\Delta_{Z}^{(m)}\int dl\,\mathrm{cos}(2m\varphi)$. Let us now label the edges of our system by $s\in\{0,...,3\}$ starting from the right edge of the sample and proceeding in counterclockwise order. In the strong-coupling regime, we find that along the $x$ ($y$) edges we have $\varphi_i=\frac{\pi}{m}(p_i+1/2)$ [$\theta_i=\frac{\pi}{m}(q_j+1/2)$] for $p_i,q_j\in\mathbb{Z}$, where $i\in\{0,2\}$, $j\in\{1,3\}$ label the respective edge. These operators satisfy $[p_i(l),q_j(l')]=\frac{im}{2\pi}\mathrm{sgn}(l-l')$. If we label the corners of a rectangular sample by $v\in\{0,...,3\}$ in counterclockwise order starting from the corner between edges 0 and 1, we can define operators acting locally on the corners as $\gamma_{2k}=e^{i\pi(p_{2k}-q_{2k+1})/m}$, $\gamma_{2k+1}=e^{i\pi(p_{2k+2}-q_{2k+1})/m}$ for $k\in\{0,1\}$. These operators commute with the Hamiltonian as they act on domain walls between segments gapped by competing mechanisms and satisfy $\mathbb{Z}_{2m}$ parafermionic commutation relations $\gamma_{v}\gamma_{v'}=\gamma_{v'}\gamma_{v}e^{-i\pi/m}$ for $v<v'$. As such, we find a single zero-energy $\mathbb{Z}_{2m}$ parafermion corner state per corner.

\begin{figure*}[!t]
	\includegraphics[width=0.8 \textwidth]{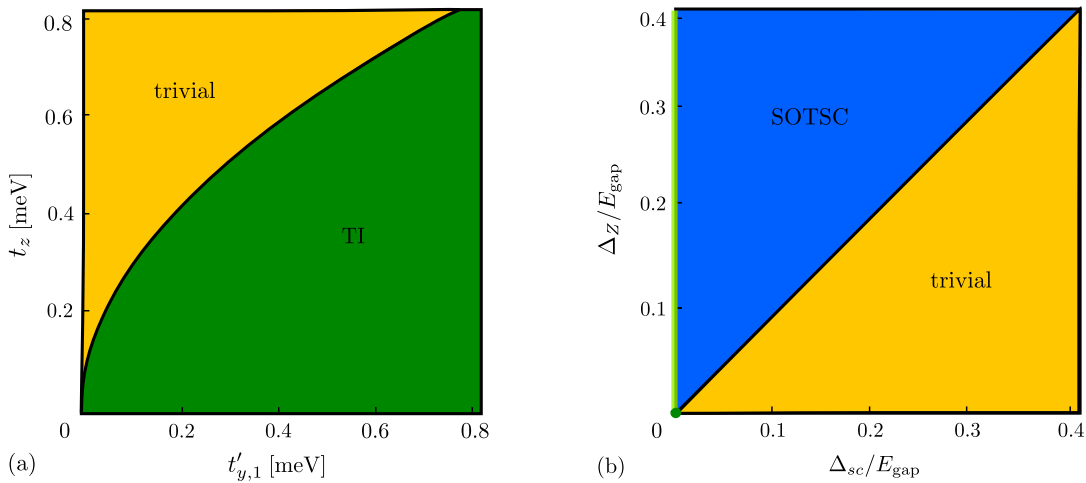}
	\caption{(a) Phase diagram of the first-order  phase (i.e., $\Delta_Z=\Delta_{sc}=0$) as a function of $t_{y,1}'$ and $t_z$.  The bulk gap closing line at $t_z=\sqrt{(t_{y,1}+t_{y,1}')(t_{y,\bar1}+t_{y,\bar1}')}$ marks the transition from the topologically trivial phase to the TI phase with a Kramers pair of gapless edge states. (b) Phase diagram of the second-order phase as a function of $\Delta_Z$ and $\Delta_{sc}$. Importantly, we consider both $\Delta_{sc}$ and $\Delta_Z$ to be smaller than the bulk gap $E_{\mathrm{gap}}$, such that the bulk gap is never closed by these terms. However, the edge gap closes and reopens for $|\Delta_{sc}|=|\Delta_Z|$, corresponding to the phase transition between the topologically trivial phase and the second-order TSC (SOTSC) phase with one Majorana corner state per corner. The green dot at $\Delta_Z=\Delta_{sc}=0$ corresponds to the first-order phase, while the light green line for $\Delta_{sc}=0$ and $\Delta_Z>0$ indicates the phase with partially gapped edge states (gapped along the $x$ direction but gapless along the $y$ direction). The other parameters are the same as in Fig.~\ref{fig:majoranas}.}
	\label{fig:phasediagram}
\end{figure*}

\section{Conclusions}
\label{sec:conclusions}

We have considered a model based on two coupled sheets of bilayer graphene in the strong SOI regime. Electrostatic confinement is used to create effective 1D quantum wires, which are then tunnel-coupled in various ways. For a certain range of parameters, this system can be brought into a topological insulator phase characterized by the presence of a Kramers pair of gapless helical edge states. Furthermore, a small in-plane magnetic field and weak proximity-induced superconductivity drive the system into a second-order topological superconducting phase with zero-energy Majorana corner states at all four corners of a rectangular sample. Even more interestingly, the fact that we are dealing with effective 1D systems allows us to take into account the effects of strong electron-electron interactions in an analytically tractable way. Using a bosonization approach, we have shown that for sufficiently strong electron-electron interactions and suitable values of chemical potential, the system can be brought into a fractional topological insulator phase with fractionally charged gapless helical edge states as well as into a fractional second-order topological superconducting phase hosting exotic $\mathbb{Z}_{2m}$ parafermion corner states, where $m$ is an odd integer determined by the position of the chemical potential.

In particular, we envision the strong SOI in the graphene bilayers to be induced by proximity to a few layers of $\mathrm{NbSe}_2$, which at the same time also induce superconductivity into the system. The recent progress in fabricating van der Waals heterostructures puts such a setup well into experimental reach. From a more general perspective, we therefore believe that our system demonstrates the potential use of electrostatically generated arrays of effective quantum wires in bilayer graphene as designer platforms to realize topologically non-trivial physics.

On the other hand, we note that while gated bilayer graphene turns out to be a particularly convenient platform to realize the model proposed here, our results can readily be adapted to different realizations of coupled 1D wires with similar low-energy properties.

\acknowledgements

This work was supported by the Swiss National Science Foundation and NCCR QSIT. This project received funding from the European Union’s Horizon 2020 research and innovation program (ERC Starting Grant, grant agreement No 757725).

\appendix

\section{Alternative realization of the second-order phase}
\label{app:mu}

In this Appendix, we comment on an alternative realization of the second-order TSC phase. For this, we tune the chemical potential to $\mu_\tau=V_0/(2\sqrt{2})+\tau\alpha$ instead of the values chosen in the main text. By replacing $\kappa\tau\rightarrow\tilde\kappa$, we note that the sector $\tilde\kappa=1$ now once again corresponds to the modes with small Fermi momenta close to $k_x=0$. Indeed, the spectrum for the sector $\tilde\kappa=1$ is identical to the one shown in Fig.~\ref{fig:spectrum_integer} and can be analyzed in the same way as in the main text. However, the sector $\tilde\kappa=\bar1$ has to be treated differently in this case. Indeed, the interlayer hopping term [see Eq.~(\ref{eq:Hzlin})] is not able to open a gap for the sector $\tilde\kappa=\bar1$ anymore, as it couples right-moving (left-moving) with right-moving (left-moving) modes. Therefore, the first-order topological insulator phase is not present in this case, as the bulk of the system is not fully gapped out. However, once superconductivity is taken into account, the gapless modes in the sector $\tilde\kappa=\bar1$ are trivially gapped out by superconductivity, while the sector $\tilde\kappa=1$ can be treated in the exact same way as before. Therefore, we find a second-order TSC phase with the same properties as in the main text.

\section{Phase diagram in the non-interacting case}
\label{app:phasediagram}

In the main text, we assumed $t_{y,1}=t_{y,\bar{1}}'=0$ and $t_{y,1}'=t_{y,\bar1}$ for analytical simplicity. However, we argued that the topological properties of the system stay qualitatively the same in an extended region of parameter space. In this Appendix, we confirm this statement by calculating a condition for the closing of the bulk gap for a more general choice of hopping amplitudes. We start by considering the first-order phase with $\Delta_{sc}=\Delta_Z=0$. To simplify matters, we focus on the case of $t_{y,\tau},t_{y,\tau}',t_z\geq0$, but our analysis can easily be extended to account for negative values of the hopping amplitudes as well. The bulk Hamiltonian is obtained from Eq.~(\ref{eq:Hky}) upon replacing $-i\partial_x\rightarrow k_x$, and has time-reversal symmetry expressed by $\mathcal{T}=i\sigma_y\rho_x\mathcal{K}$, where $\mathcal{K}$ denotes the complex conjugation. Therefore, our system belongs to the symmetry class AII~\cite{Ryu2010}.
As long as $t_{y,1}',t_{y,\bar1}>t_{y,1},t_{y,\bar1}'$,  the bulk gap can only close at $k_x=k_y=0$, where the eigenenergies are explicitly given by
\begin{align}
E_{1,\pm}\hspace{3mm}&=\pm t_z,\\
E_{2,\pm,\pm}&=\pm\Big(t_{y,1}+t_{y,\bar1}+t_{y,1}'+t_{y,\bar1}'\nonumber\\&\hspace{4mm}\pm\sqrt{(t_{y,1}-t_{y,\bar1}+t_{y,1}'-t_{y,\bar1}')^2+4t_z^2}\Big)/2.
\end{align}
%
%
%
Thus, we find that the bulk of the system is fully gapped except for the values $t_z=0$ or $t_z^2=(t_{y,1}+t_{y,1}')(t_{y,\bar1}+t_{y,\bar1}')$.
These conditions define the potential boundaries between topologically non-equivalent phases. In our case, we can identify the region corresponding to the topologically non-trivial phase by checking for the existence of gapless edge states: In the main text, we argued that the system hosts a Kramers pair of gapless edge states for $t_{y,1}=t_{y,\bar1}'=0$ and $t_{y,\bar1}=t_{y,1}'\gg t_z>0$. Therefore, we identify the region of the phase diagram for which $0<t_z^2<(t_{y,1}+t_{y,1}')(t_{y,\bar1}+t_{y,\bar1}')$ as the topologically non-trivial one. This is visualized in Fig.~\ref{fig:phasediagram}(a).

Let us now turn to the second-order phase. In the presence of a magnetic field, $\Delta_Z\neq0$, time-reversal symmetry is broken, while superconductivity, $\Delta_{sc}\neq 0$, enforces particle-hole symmetry. This places our system in the symmetry class D~\cite{Ryu2010}. Assuming that $\Delta_{sc}$ and $\Delta_Z$ are much smaller than the bulk gap, these terms cannot result in a closing of the bulk gap. However, we find that the edge gap closes at the points $|\Delta_{sc}|=|\Delta_{Z}|$, which separates the topologically non-trivial second-order phase with one zero-energy Majorana corner state per corner from the trivial one with no corner states. The corresponding phase diagram is shown in Fig.~\ref{fig:phasediagram}(b).

\bibliographystyle{unsrt}

\begin{thebibliography}{}
	
	\bibitem{Hasan2010}
	M. Z. Hasan and C. L. Kane, Rev. Mod. Phys. {\bf 82}, 3045 (2010).
	\bibitem{Qi2011}
	X.-L. Qi and S.-C. Zhang, Rev. Mod. Phys. {\bf 83}, 1057 (2011).
	\bibitem{Sato2017}
	M. Sato and Y. Ando, Rep. Prog. Phys. {\bf 80}, 076501 (2017).
	
	\bibitem{Benalcazar2014}
	W. A. Benalcazar, J. C. Y. Teo, and T. L. Hughes, Phys. Rev. B {\bf 89}, 224503 (2014).
	\bibitem{Benalcazar2017}
	W. A. Benalcazar, B. A. Bernevig, and T. L. Hughes, Science {\bf 357}, 61 (2017).
	\bibitem{Benalcazar2017b}
	W. A. Benalcazar, B. A. Bernevig, and T. L. Hughes, Phys. Rev. B {\bf 96}, 245115 (2017).
	\bibitem{Song2017}
	Z. Song, Z. Fang, and C. Fang, Phys. Rev. Lett. {\bf 119}, 246402 (2017).
	\bibitem{Peng2017}
	Y. Peng, Y. Bao, and F. von Oppen, Phys. Rev. B {\bf 95}, 235143 (2017).
	\bibitem{Imhof2017}
	S. Imhof, C. Berger, F. Bayer, H. Brehm, L. Molenkamp, T. Kiessling, F. Schindler, C. H. Lee, M. Greiter, T. Neupert, and R. Thomale, Nat. Phys. {\bf 14}, 925 (2018).
	\bibitem{Geier2018}
	M. Geier, L. Trifunovic, M. Hoskam, and P. W. Brouwer, Phys. Rev. B {\bf 97}, 205135 (2018).
	\bibitem{Schindler2018}
	F. Schindler, A. M. Cook, M. G. Verginory, Z. Wang, S. S. P. Parking, B. A. Bernevig, and T. Neupert, Science Adv. {\bf 4}, 6 (2018).
	\bibitem{Hsu2018}
	C.-H. Hsu, P. Stano, J. Klinovaja, and D. Loss, Phys. Rev. Lett. {\bf 121}, 196801 (2018).
	\bibitem{Ezawa2018b}
	M. Ezawa, Sci. Rep. {\bf 9}, 5286 (2019).
	\bibitem{Ezawa2018c}
	M. Ezawa, Phys. Rev. Lett. {\bf 121}, 116801 (2018).
	\bibitem{Zhu2018}
	X. Zhu, Phys. Rev. B {\bf 97}, 205134 (2018).
	\bibitem{Wang2018}
	Q. Wang, C.-C. Liu, Y.-M. Lu, and F. Zhang, Phys. Rev. Lett. {\bf 121}, 186801 (2018).
	\bibitem{Yan2018}
	Z. Yan, F. Song, and Z. Wang, Phys. Rev. Lett. {\bf 121}, 096803 (2018).
	\bibitem{Liu2018}
	T. Liu, J. J. He, and F. Nori, Phys. Rev. B {\bf 98}, 245413 (2018).
	\bibitem{Zhang2018}
	X. Zhang, H.-X. Wang, Z.-K. Lin, Y. Tian, B. Xie, M.-H. Lu, Y.-F. Chen, and J.-H. Jiang, Nat. Phys. {\bf 15}, 582 (2019).
	\bibitem{Wang2018b}
	Q. Wang, D. Wang, and Q.-H. Wang, EPL {\bf 124}, 50005 (2018).
	\bibitem{Volpez2018}
	Y. Volpez, D. Loss, and J. Klinovaja, Phys. Rev. Lett. {\bf 122}, 126402 (2019).
	\bibitem{Plekhanov2019}
	K. Plekhanov, M. Thakurathi, D. Loss, and J. Klinovaja, Phys. Rev. Res. {\bf 1}, 032013(R) (2019).
	\bibitem{Calugaru2019}
	D. Calugaru, V. Juricic, and B. Roy, Phys. Rev. B {\bf 99}, 041301(R) (2019).
	\bibitem{Agarwala2019}
	A. Agarwala, V. Juricic, and B. Roy, arXiv:1902.00507.
	\bibitem{Yan2019}
	Z. Yan, Phys. Rev. Lett. {\bf 123}, 177001 (2019).
	\bibitem{Franca2019}
	S. Franca, D. V. Efremov, and I. C. Fulga, Phys. Rev. B {\bf 100}, 075415 (2019).
	\bibitem{Zhang2019a}
	R.-X. Zhang, W. S. Cole, and S. Das Sarma, Phys. Rev. Lett. {\bf 122}, 187001 (2019).
	\bibitem{Zhang2019b}
	S.-B. Zhang and B. Trauzettel, arXiv:1905.09308.
	
	\bibitem{You2018a}
	Y. You, D. Litinski, and F. von Oppen, Phys. Rev. B {\bf 100}, 054513 (2019).
	\bibitem{You2018b}
	Y. You, T. Devakul, F. J. Burnell, and T. Neupert, Phys. Rev. B {\bf 98}, 235102 (2018).
	\bibitem{Laubscher2019}
	K. Laubscher, D. Loss, and J. Klinovaja, Phys. Rev. Res. {\bf 1}, 032017(R) (2019).
	
	\bibitem{CastroNeto2009}
	A. H. Castro Neto, F. Guinea, N. M. R. Peres, K. S. Novoselov, and A. K. Geim, Rev. Mod. Phys. {\bf 81}, 109 (2009).
	\bibitem{DasSarma2011}
	S. Das Sarma, S. Adam, E. H. Hwang, and E. Rossi, Rev. Mod. Phys. {\bf 83}, 407 (2011).
	\bibitem{Kane2005}
	C. L. Kane and E. J. Mele, Phys. Rev. Lett. {\bf 95}, 226801 (2005).
	
	\bibitem{Kiesel2012}
	M. L. Kiesel, C. Platt, W. Hanke, D. A. Abanin, and R. Thomale, Phys. Rev. B {\bf 86}, 020507(R) (2012).
	\bibitem{Klinovaja2012c}
	J. Klinovaja, S. Gangadharaiah, and D. Loss, Phys. Rev. Lett. {\bf 108}, 196804 (2012).
	\bibitem{Sau2013}
	J. D. Sau and S. Tewari, Phys. Rev. B {\bf 88}, 054503 (2013).
	\bibitem{Egger2012}
	R. Egger and K. Flensberg, Phys. Rev. B {\bf 85}, 235462 (2012).
	\bibitem{Jose2015}
	P. San-Jose, J. L. Lado, R. Aguado, F. Guinea, and J. Fern\'{a}ndez-Rossier, Phys. Rev. X {\bf 5}, 041042 (2015).
	\bibitem{Schaffer2012}
	A. M. Black-Schaffer, Phys. Rev. Lett. {\bf 109}, 197001 (2012).
	\bibitem{Dutreix2014}
	C. Dutreix, M. Guigou, D. Chevallier, and C. Bena, Eur. Phys. J. B {\bf 87}, 296 (2014).
	\bibitem{Klinovaja2013}
	J. Klinovaja and D. Loss, Phys. Rev. X {\bf 3}, 011008 (2013).
	\bibitem{Marganska2018}
	M. Marganska, L. Milz, W. Izumida, C. Strunk, and M. Grifoni, Phys. Rev. B {\bf 97}, 075141 (2018).
	
	\bibitem{Gmitra2009}
	M. Gmitra, S. Konschuh, C. Ertler, C. Ambrosch-Draxl, and J. Fabian, Phys. Rev. B {\bf 80}, 235431 (2009).
	
	\bibitem{Kaloni2014}
	T. P. Kaloni, L. Kou, T. Frauenheim, and U. Schwingenschl\"{o}gl, Appl. Phys. Lett. {\bf 105}, 233112 (2014).
	\bibitem{Avsar2014}
	A. Avsar, J. Y. Tan, T. Taychatanapat, J. Balakrishnan, G. K. W. Koon, Y. Yeo, J. Lahiri, A. Carvalho, A. S. Rodin, E. C. T. O'Farrell, G. Eda, A. H. Castro Neto, and B. \"{O}zyilmaz, Nat. Commun. {\bf 5}, 4875 (2014).
	\bibitem{Gmitra2015}
	M. Gmitra and J. Fabian, Phys. Rev. B {\bf 92}, 155403 (2015).
	\bibitem{Wang2015}
	Z. Wang, D.-K. Ki, H. Chen, H. Berger, A. H. MacDonald, and A. F. Morpurgo, Nat. Commun. {\bf 6}, 8339 (2015).
	\bibitem{Gmitra2016}
	M. Gmitra, D. Kochan, P. H\"{o}gl, and J. Fabian, Phys. Rev. B {\bf 93}, 155104 (2016).
	\bibitem{Wang2016}
	Z. Wang, D.-K. Ki, J. Y. Khoo, D. Mauro, H. Berger, L. S. Levitov, and A. F. Morpurgo, Phys. Rev. X {\bf 6}, 041020 (2016).
	\bibitem{Alsharari2016}
	A. M. Alsharari, M. M. Asmar, and S. E. Ulloa, Phys. Rev. B {\bf 94}, 241106(R) (2016).
	\bibitem{Kochan2017}
	D. Kochan, S. Irmer, and J. Fabian, Phys. Rev. B {\bf 95}, 165415 (2017).
	\bibitem{Gani2019}
	Y. S. Gani, H. Steinberg, and E. Rossi, Phys. Rev. B {\bf 99}, 235404 (2019).
	\bibitem{Gmitra2017}
	M. Gmitra and J. Fabian, Phys. Rev. Lett. {\bf 119}, 146401 (2017).
	\bibitem{Khoo2017}
	J. Y. Khoo, A. F. Morpurgo, and L. Levitov, Nano Lett. {\bf 17}, 7003 (2017).
	\bibitem{Zihlmann2018}
	S. Zihlmann, A. W. Cummings, J. H. Garcia, M. Kedves, K. Watanabe, T. Taniguchi, C. Sch\"{o}nenberger, and P. Makk, Phys. Rev. B {\bf 97}, 075434 (2018).
	
	\bibitem{Martin2008}
	I. Martin, Y. M. Blanter, and A. F. Morpurgo, Phys. Rev. Lett. {\bf 100}, 036804 (2008).
	\bibitem{Li2016}
	J. Li, K. Wang, K. J. McFaul, Z. Zern, Y. Ren, K. Watanabe, T. Taniguchi, Z. Qiao, and J. Zhu, Nat. Nanotechnol. {\bf 11}, 1060 (2016).
	\bibitem{Yin2016}
	L.-J. Yin, H. Jiang, J.-B. Qiao, and L. He, Nat. Commun. {\bf 7}, 11760 (2016).
	\bibitem{Ju2015}
	L. Ju, Z. Shi, N. Nair, Y. Lv, C. Jin, J. Velasco Jr, C. Ojeda-Aristizabal, H. A. Bechtel, M. C. Martin, A. Zettl, J. Analytis, and F. Wang, Nature (London) {\bf 520}, 650 (2015).
	\bibitem{Klinovaja2012a}
	J. Klinovaja, G. J. Ferreira, and D. Loss, 	Phys. Rev. B {\bf 86}, 235416 (2012).
	
	\bibitem{Poilblanc1987}
	D. Poilblanc, G. Montambaux, M. H\'{e}ritier, and P. Lederer, Phys. Rev. Lett. {\bf 58}, 270 (1987).
	\bibitem{Gorkov1995}
	L. P. Gorkov and A. G. Lebed, Phys. Rev. B {\bf 51}, 3285 (1995).
	\bibitem{Kane2002}
	C. L. Kane, R. Mukhopadhyay, and T. C. Lubensky, Phys. Rev. Lett. {\bf 88}, 036401 (2002).
	\bibitem{Teo2014}
	J. C. Y. Teo and C. L. Kane, Phys. Rev. B {\bf 89}, 085101 (2014).
	\bibitem{Klinovaja2014a}
	J. Klinovaja and Y. Tserkovnyak, Phys. Rev. B {\bf 90}, 115426 (2014).
	\bibitem{Klinovaja2014b}
	J. Klinovaja and D. Loss, Eur. Phys. J. B {\bf 87}, 171 (2014).
	\bibitem{Sagi2014}
	E. Sagi and Y. Oreg, Phys. Rev. B {\bf 90}, 201102(R) (2014).
	\bibitem{Neupert2014}
	T. Neupert, C. Chamon, C. Mudry, and R. Thomale, Phys. Rev. B {\bf 90}, 205101 (2014).
	\bibitem{Klinovaja2015}
	J. Klinovaja, Y. Tserkovnyak, and D. Loss, Phys. Rev. B {\bf 91}, 085426 (2015).
	\bibitem{Sagi2015}
	E. Sagi and Y. Oreg, Phys. Rev. B {\bf 92}, 195137 (2015).
	\bibitem{Meng2015b}
	T. Meng, Phys. Rev. B {\bf 92}, 115152 (2015).
	\bibitem{Teo2016}
	S. Sahoo, Z. Zhang, and J. C. Y. Teo, Phys. Rev. B {\bf 94}, 165142 (2016).

	\bibitem{Ando2000}
	T. Ando, J. Phys. Soc. Jpn. {\bf 69}, 1757 (2000).
	\bibitem{Hernando2006}
	D. Huertas-Hernando, F. Guinea, and A. Brataas, Phys. Rev. B {\bf 74}, 155426 (2006).
	\bibitem{Kuemmeth2008}
	F. Kuemmeth, S. Ilani, D. C. Ralph, and P. L. McEuen, Nature (London) {\bf 452}, 448 (2008).
	\bibitem{Izumida2009}
	W. Izumida, K. Sato, and R. Saito, J. Phys. Soc. Jpn. {\bf 78}, 074707 (2009).
	\bibitem{Chico2009}
	L. Chico, M. P. L\'{o}pez-Sancho, and M. C. Mu\~{n}oz, Phys. Rev. B {\bf 79}, 235423 (2009).
	\bibitem{Klinovaja2011a}
	J. Klinovaja, M. J. Schmidt, B. Braunecker, and D. Loss, Phys. Rev. Lett. {\bf 106}, 156809 (2011).
	\bibitem{Klinovaja2011b}
	J. Klinovaja, M. J. Schmidt, B. Braunecker, and D. Loss, Phys. Rev. B {\bf 84}, 085452 (2011).
	\bibitem{Steele2013}
	G. A. Steele, F. Pei, E. A. Laird, J. M. Jol, H. B. Meerwaldt, and L. P. Kouwenhoven, Nat. Commun. {\bf 4}, 1573 (2013).
	
	\bibitem{soi}
	Note that we did not include the intrinsic Kane-Mele SOI as well as a staggered sublattice potential term which is in principle allowed by symmetry, as those terms have been predicted to be orders of magnitude smaller than the SOI induced by proximity to the TMD layers~\cite{Gmitra2009,Wang2015,Wang2016}.
	\bibitem{Klinovaja2012b}
	J. Klinovaja and D. Loss, Phys. Rev. B {\bf 86}, 085408 (2012).
	\bibitem{Giamarchi2004}
	T. Giamarchi, \textit{Quantum Physics in One Dimension} (Oxford University Press, Oxford, 2004).
	
	\bibitem{Cheng2012}
	M. Cheng, Phys. Rev. B {\bf 86}, 195126 (2012).
	\bibitem{Clarke2013}
	D. J. Clarke, J. Alicea and K. Shtengel, Nat. Commun. {\bf 4}, 1348 (2013).
	\bibitem{Motruk2013}
	J. Motruk, E. Berg, A. M. Turner, and F. Pollmann, Phys. Rev. B {\bf 88}, 085115 (2013).
	\bibitem{Lindner2012}
	N. H. Lindner, E. Berg, G. Refael, and A. Stern, Phys. Rev. X {\bf 2}, 041002 (2012).
	
	
	\bibitem{Ryu2010}
	S. Ryu, A. P. Schnyder, A. Furusaki, and A. W. W. Ludwig, New J. Phys. {\bf 12}, 065010 (2010).
	
	
\end{thebibliography}

\end{document}